\documentclass[usenatbib]{mn2e}
\usepackage{graphicx}
\usepackage{amsmath}
\usepackage{amssymb}
\usepackage{color}
\usepackage{url}

\usepackage{threeparttable}

\newcommand\lsim{\mathrel{\rlap{\lower4pt\hbox{\hskip1pt$\sim$}}
        \raise1pt\hbox{$<$}}}
\newcommand\gsim{\mathrel{\rlap{\lower4pt\hbox{\hskip1pt$\sim$}}
        \raise1pt\hbox{$>$}}}
\newcommand{\lya}{\ifmmode\mathrm{Ly}\alpha\else{}Ly$\alpha$\fi}
\newcommand{\lyb}{\ifmmode\mathrm{Ly}\beta\else{}Ly$\beta$\fi}
\newcommand{\igm}{\ifmmode\mathrm{IGM}\else{}IGM\fi}
\newcommand{\lae}{\ifmmode\mathrm{LAE}\else{}LAE\fi}
\newcommand{\h}{\ifmmode\mathrm{H}\else{}H\fi}
\newcommand{\hi}{\ifmmode\mathrm{H\,{\scriptscriptstyle I}}\else{}H\,{\scriptsize I}\fi}
\newcommand{\hii}{\ifmmode\mathrm{H\,{\scriptscriptstyle II}}\else{}H\,{\scriptsize II}\fi}
\newcommand{\cmb}{\ifmmode\mathrm{CMB}\else{}CMB\fi}
\newcommand{\qso}{\ifmmode\mathrm{QSO}\else{}QSO\fi}
\newcommand{\eor}{\ifmmode\mathrm{EoR}\else{}EoR\fi}
\newcommand{\heii}{\ifmmode\mathrm{He\,{\scriptscriptstyle II}}\else{}He\,{\scriptsize II}\fi}
\newcommand{\heiii}{\ifmmode\mathrm{He\,{\scriptscriptstyle III}}\else{}He\,{\scriptsize III}\fi}
\newcommand{\ciii}{\ifmmode\mathrm{C\,{\scriptscriptstyle III]}}\else{}C\,{\scriptsize III]}\fi}
\newcommand{\oiii}{\ifmmode\mathrm{O\,{\scriptscriptstyle III}}\else{}O\,{\scriptsize III}\fi}
\newcommand{\aliii}{\ifmmode\mathrm{Al\,{\scriptscriptstyle III}}\else{}Al\,{\scriptsize III}\fi}
\newcommand{\mgii}{\ifmmode\mathrm{Mg\,{\scriptscriptstyle II}}\else{}Mg\,{\scriptsize II}\fi}
\newcommand{\fe}{\ifmmode\mathrm{Fe}\else{}Fe\fi}
\newcommand{\nv}{\ifmmode\mathrm{N\,{\scriptscriptstyle V}}\else{}N\,{\scriptsize V}\fi}
\newcommand{\niv}{\ifmmode\mathrm{N\,{\scriptscriptstyle IV]}}\else{}N\,{\scriptsize IV]}\fi}
\newcommand{\cii}{\ifmmode\mathrm{C\,{\scriptscriptstyle II}}\else{}C\,{\scriptsize II}\fi}
\newcommand{\civ}{\ifmmode\mathrm{C\,{\scriptscriptstyle IV}}\else{}C\,{\scriptsize IV}\fi}
\newcommand{\siv}{\ifmmode\mathrm{Si\,{\scriptscriptstyle IV}}\else{}Si\,{\scriptsize IV}\fi}
\newcommand{\siii}{\ifmmode\mathrm{Si\,{\scriptscriptstyle II}}\else{}Si\,{\scriptsize II}\fi}
\newcommand{\siiii}{\ifmmode\mathrm{Si\,{\scriptscriptstyle III]}}\else{}Si\,{\scriptsize III]}\fi}
\newcommand{\ovi}{\ifmmode\mathrm{O\,{\scriptscriptstyle VI}}\else{}O\,{\scriptsize VI}\fi}
\newcommand{\sioiv}{\ifmmode\mathrm{Si\,{\scriptscriptstyle IV}\,\plus O\,{\scriptscriptstyle IV]}}\else{}Si\,{\scriptsize IV}\,+O\,{\scriptsize IV]}\fi}
\newcommand{\Msun}{M_\odot}

\newcommand{\nf}{x_{\rm HI}}
\newcommand{\avenf}{$\bar{x}_{\rm HI}$}
\newcommand{\faintgal}{\textsc{\small Faint galaxies}}
\newcommand{\brightgal}{\textsc{\small Bright galaxies}}
\newcommand{\smallHII}{\textsc{\small Small \hii{}}}
\newcommand{\largeHII}{\textsc{\small Large \hii{}}}

\pdfoutput=1
\begin{document}

\title[Reionisation from J1120+0641]{Are we witnessing the epoch of reionisation at $\bmath{z=7.1}$ from the spectrum of J1120+0641?}

\author[B. Greig et al.] {Bradley~Greig$^{1}$\thanks{E-mail:~bradley.greig@sns.it}, Andrei~Mesinger$^{1}$, Zolt{\'a}n~Haiman$^{2}$, \& Robert~A.~Simcoe$^{3,4}$\\
$^1$Scuola Normale Superiore, Piazza dei Cavalieri 7, I-56126 Pisa, Italy \\
$^2$Department of Astronomy, Columbia University, 550 West 120th Street, New York, NY 10027, USA \\
$^3$Massachusetts Institute of Technology, 77 Massachusetts Ave, Cambridge, MA 02139, USA \\
$^4$MIT-Kavli Center for Astrophysics and Space Research \\
}

\maketitle \begin{abstract}
\noindent
We quantify the presence of \lya\ damping wing absorption from a partially-neutral intergalactic medium (IGM) in the spectrum of the $z=7.08$ QSO, ULASJ1120+0641. Using a Bayesian framework, we simultaneously account for uncertainties in: (i) the intrinsic QSO emission spectrum; and (ii) the distribution of cosmic \hi{} patches during the epoch of reionisation (EoR).  For (i) we use a new intrinsic \lya{} emission line reconstruction method (Greig et al.), sampling a covariance matrix of emission line properties built from a large database of moderate-$z$ QSOs.  For (ii), we use the Evolution of 21-cm Structure (EOS; Mesinger et al.) simulations, which span a range of physically-motivated EoR models. We find strong evidence for the presence of damping wing absorption redward of \lya\ (where there is no contamination from the \lya\ forest). Our analysis implies that the EoR is not yet complete by $z=7.1$, with the volume-weighted IGM neutral fraction constrained to $\bar{x}_{\hi{}} = 0.40\substack{+0.21 \\ -0.19}$ at $1\sigma$ ($\bar{x}_{\hi{}} = 0.40\substack{+0.41 \\ -0.32}$ at $2\sigma$).  This result is insensitive to the EoR morphology.  Our detection of significant neutral \hi{} in the IGM at $z=7.1$ is consistent with the latest {\it Planck} 2016 measurements of the CMB Thompson scattering optical depth (Planck Collaboration XLVII).
\end{abstract} 

\section{Introduction}

The epoch of reionisation (\eor{}) signals the end of the cosmic dark ages, when ionising radiation from the first stars and galaxies spreads throughout the Universe, beginning the last major baryonic phase change. This \eor{} is rich in astrophysical information, providing insights into the formation, properties and evolution of the first cosmic structures in the Universe.

Several recent $z \gsim 6$ observations have provided (controversial) information about the EoR (for a review, see eg., \citealt{Mesinger:2016book}). These come either from integral constraints on \hii{} provided by the Thomson scattering of \cmb{} photons \citep[e.g.][]{Collaboration:2015p4320,George:2015p5869}, or \lya\ absorption by putative cosmic \hi{} patches along the lines of sight towards $z \gsim 6$ objects.  Since the cross-section at the \lya\ line centre is large enough to saturate transmission even in the ionised intergalactic medium (IGM; requiring only trace values of neutral hydrogen: $\nf \gsim 10^{-4}-10^{-5}$), the latter constraints generally rely on the damping wing of the line.  The relative flatness of the damping wing with frequency contributes a smooth absorption profile, which can result in optical depths of order a few during the EoR at frequencies around the redshifted \lya\ line.

For galaxies, constraining damping wing absorption must be done with large samples, using their redshift evolution and/or clustering properties \citep[e.g.][]{HaimanSpaans1999,Ouchi:2010p1,Stark:2010p1,Pentericci:2011p1,Ono:2012p1,Caruana:2014p1,Schenker:2014p1}. QSOs however can be much brighter, allowing the detection of the EoR damping wing from a single spectrum. Most bright $z\gsim6$ QSOs have a large region of detectable flux blueward of the rest frame 1216 \AA, where the flux from the QSO itself is thought to facilitate transmission even for photons redshifting into the \lya\ resonant core. If this so-called near zone is large, then the imprint of an EoR damping wing can be isolated as a smooth absorption component on top of the fluctuating resonant absorption (the \lya\ forest) inside the near zone \citep[e.g.][]{Mesinger:2004p5737,Bolton:2007p5879,Bolton:2007p3623,Schroeder:2013p919}. The intrinsic (unabsorbed) \lya{} line profile can be reconstructed from the red side of the line (where absorption is minimal), assuming the line is symmetric (a reasonable assumption for bright QSOs; e.g. \citealt{Kramer:2009p920}). However, one is faced with the challenge of modelling the statistics of the \lya\ forest in the near zone, which depend on the ionising background and temperature. 

The highest redshift QSO observed to date, the $z\sim7.1$ object ULASJ1120+0641 \citep[][hereafter ULASJ1120]{Mortlock:2011p1049}, appears to have an uncharacteristically-small near zone: $\sim2$~proper Mpc, roughly a factor of 3--4 smaller than generally found in $z\sim6$ QSOs of comparable brightness \citep{Carilli:2010p1}.\footnote{Note that quantitative EoR constraints using just the apparent size of the near zone require assumptions about the QSO age, environment, and ionisation history (e.g. \citealt{Mesinger:2004p5737, Maselli:2007p5744, Bolton:2011p1063}).}  If the IGM is indeed undergoing reionisation at $z=7.1$, cosmic \hi{} patches along the line of sight could be close enough to the quasar to imprint a detectable damping wing signature away from the near zone edge, even redward of the \lya\ emission line centre.\footnote{In principle, strong \lya{} attenuation could also be caused by a damped \lya{} absorber (DLA) intersected along the line of sight.  However, such a system would have to have both an extremely high column density (${\rm log_{10}}(N_{\hi{}}/{\rm cm^{-2}}) = 20.5 - 21$; \citealt{Simcoe:2012p1057,Schroeder:2013p919}), and an extremely low metallicity ($Z \lsim10^{-4} Z_\odot$; e.g. \citealt{Simcoe:2012p1057,Maio:2013p1}). As we discuss further below, the number density of such objects in a random IGM sightline is extremely small \citep[e.g.][]{Prochaska:2009p3339}.}
Redward of the \lya\ line centre (and redward of any redshifted cosmological infall; e.g. \citealt{Barkana:2004p4778}), there is no contribution from resonant absorption in the \lya\ forest (see Section~\ref{sec:Constraints}).  Not having to model the \lya\ forest simplifies the analysis considerably.   However, as it might contain a damping wing imprint, the red side of the observed emission line can no longer be used to independently reconstruct the \lya\ line profile.   Moreover, the intrinsic \lya\ emission profile can vary significantly from object to object, complicating the usefulness of a composite emission template.  Thus, {\it any analysis of the IGM damping wing would need to fold-in the significant uncertainties in the shape and amplitude of the intrinsic \lya{} line profile.}

The difficulty in reconstructing the intrinsic emission of ULASJ1120 is further exacerbated by its peculiar emission line features. \citet{Mortlock:2011p1049} report an extremely large \civ{} blue-shift relative to its systemic redshift, which is larger than what has been observed in 99.9 per cent of all known QSOs. \citet{Bosman:2015p5005} recently suggested that the observed \lya\ emission of ULASJ1120 might be consistent with the subsample of objects with similar \civ{} properties, potentially alleviating the need for additional damping wing absorption.\footnote{
  Note that the sample of SDSS QSOs used in the \citet{Bosman:2015p5005} analysis are selected purely on their \civ{} equivalent width and the velocity offset of the \civ{} emission line relative to the \civ{} emission line of ULASJ1120+0641. Therefore, they are not necessarily selecting anomalous QSOs like ULASJ1120. Furthermore, these SDSS QSOs use only the \ciii{} line for the redshift determination. The \ciii{} line complex is a blend of several emission lines (\aliii{}, \siiii{} and \ciii{}) and as a result the contributions from the different individual line profiles may lead to small biases in their determined velocity offsets.
}  Correctly accounting for correlations of line properties is therefore critical for any robust claims on reionisation from ULASJ1120.

In this work, we re-analyse the spectrum of ULASJ1120, improving on prior work with a combination of the following:
\begin{itemize}
\item We use the recently developed intrinsic \lya{} emission line reconstruction method \citep{Greig:2016p1}, which samples a covariance matrix of emission line properties from $\sim1500$ moderate-$z$ unobscured QSOs.
\item We use the latest, large-scale (1.6 Gpc on a side), physics-rich simulations of the EoR \citep{Mesinger:2016p1} to extract $10^5$ sightlines of opacity from $\sim 10^{12} \Msun$ halos (typical of bright QSOs; e.g. \citealt{Fan:2006p4005, Mortlock:2011p1049}).
\item We fold all uncertainties into a Bayesian framework, recovering robust constraints on the IGM neutral fraction at $z=7.1$, which, for the first time, include rigorous statistical confidence intervals.
\end{itemize}

The remainder of this paper is organised as follows. In Section~\ref{sec:Reconstruction} we summarise the key components of the intrinsic \lya{} reconstruction method, the observed spectrum of ULASJ1120 to be used in this analysis and the recovery of the reconstructed \lya{} line profile. In Section~\ref{sec:IGMDampingWing} we discuss the semi-numerical reionisation simulations and the extraction of the synthetic damping wing profiles and in Section~\ref{sec:Constraints} we outline our combined analysis. In Section~\ref{sec:results} we discuss the constraints on the IGM neutral fraction resulting from the imprint of the IGM damping wing, and in Section~\ref{sec:Discussion} we consider the possibility of a DLA contributing the damping wing imprint. Finally, in Section~\ref{sec:Conclusion} we finish with our closing remarks. Throughout this work, we adopt the background cosmological parameters: ($\Omega_\Lambda$, $\Omega_{\rm M}$, $\Omega_b$, $n$, $\sigma_8$, $H_0$) = (0.69, 0.31, 0.048, 0.97, 0.81, 68 km s$^{-1}$ Mpc$^{-1}$), consistent with cosmic microwave background anisotropy measurements by the Planck satellite \citep{Collaboration:2015p4320} and unless otherwise stated, distances are quoted in comoving units.

\section{Method}

\subsection{Reconstruction of the intrinsic \lya{} profile} \label{sec:Reconstruction}

The analysis of the \igm{} damping wing imprint within the spectrum of ULASJ1120 hinges on the recovery of the intrinsic \lya{} emission line profile. Already, in the previous section we have alluded to the difficulties in applying a QSO composite template to ULASJ1120. Within this work, we utilise the recently developed covariance matrix method of \citet{Greig:2016p1} to obtain our reconstructed estimate of the intrinsic \lya{} emission line profile.\footnote{An alternative approach could be to reconstruct a template using a principle component analysis (PCA) \citep[e.g.][]{Boroson:1992p4641,Francis:1992p5021,Suzuki:2005p5157,Suzuki:2006p4770,Lee:2011p1738,Paris:2011p4774, Simcoe:2012p1057}. However, typically this is used for characterising the mean QSO composite obtained from fits to the full QSO spectrum with the fewest eigenvectors. How well this approach would work in reconstructing the intrinsic profile of \lya{}, and characterising the properties of an individual source as peculiar as ULASJ1120, is beyond the scope of this work.}

In this work, we make use of the \citet{Simcoe:2012p1057} ULASJ1120 spectrum, obtained from the FIRE infrared spectrometer \citep{2008Simcoe} on the Magellan/Baade telescope. This spectrum offers an order of magnitude improvement in spatial (frequency) resolution compared to the \citet{Mortlock:2011p1049} discovery spectrum, with a similar signal to noise.\footnote{In order to test the robustness of our fit to ULASJ1120, we additionally performed our MCMC fitting approach on the \citet{Mortlock:2011p1049} discovery spectrum. While this spectrum has an order of magnitude lower resolution, this should not impact the recovery of the strong high-ionisation lines. Not shown here, we confirm that indeed we do recover similar fits for all emission line features.} Throughout this work, we report all results in the QSO rest-frame, with which we convert from the observed frame using the atomic [C\,{\scriptsize II}] transition. The resulting ULASJ1120 redshift is $z=7.0842\pm0.0004$ \citep{Venemans:2012p4996}\footnote{The uncertainty on the redshift determination is implicitly accounted for within our MCMC fitting approach, by allowing the velocity offset of each individual emission line to be a free parameter.}.

\subsubsection{Reconstruction procedure} \label{sec:summary}

Within this section we briefly summarise the major steps of the intrinsic \lya{} profile reconstruction method of \citet{Greig:2016p1}, and refer the reader to that work for more in-depth discussions.  Our approach is based on a covariance matrix characterising the emission line parameters.  We first select a subsample of moderate-$z$ ($2.08 < z < 2.5$), high signal to noise (S/N $>15$) QSOs from SDSS-III (BOSS) DR12 \citep{Dawson:2013p5160,Alam:2015p5162}.  Each QSO in our final sample of 1673 (using the `Good' sample, which provides tighter constraints on the reconstruction profile) is then fit with a single power-law continuum, and a set of Gaussian profiles to characterise the emission lines and any possible absorption features contaminating the observed QSO spectrum. Each Gaussian profile is described by three parameters: a line width, peak height and velocity offset from the systemic redshift.  For \lya{} and \civ{} we allow for both a broad and narrow component Gaussian to describe the line profile, and single components for all other high and low-ionisation lines.  After fitting all QSOs, and performing a visual quality assessment to refine our QSO sample, we construct our covariance matrix from the four most prominent high ionisation lines, \lya{}, \civ{}, \sioiv{} and \ciii{}.
  
With the covariance matrix in hand, we can reconstruct the intrinsic \lya{} line profile of ULASJ1120 as follows:
\begin{itemize}
\item Using the fitting procedure described above, we fit the spectrum of ULASJ1120
  far redward of the extent of the \lya{} line profile.  We chose $\lambda > 1275$\AA\ as the blue edge for the fit somewhat arbitrarily, but verify that this choice does not have an impact on the results.
\item From this fit (shown in Figure~\ref{fig:Spectrum}), we obtain estimates of ULASJ1120's continuum and of its \sioiv{}, \civ{} and \ciii{} line profiles.
\item Using these estimates,  we collapse the 18-dimensional (Gaussian distributed) covariance matrix into a six dimensional estimate of the intrinsic \lya{} emission line profile (two component Gaussian each with an amplitude, width and velocity offset).
\item We apply a prior within the range $1230 < \lambda < 1275$\AA. This is performed by simultaneously fitting the \lya\ +  \nv{} ($1240.81$\AA) and \siii{} ($1262.59$\AA) lines (where we sample the \lya{} profile from the six dimensional distribution), using the observed noise from the FIRE spectrum to obtain a $\chi^{2}$ likelihood for the reconstructed profile. In other words, we require our profiles to fit the observed spectrum over the range $1230 < \lambda < 1275$\AA.
  This final step notably reduces the errors on the reconstructed \lya{} profile (shown in Figure~\ref{fig:Profile}), ruling out extreme profiles which are inconsistent with the actual observed spectrum.  We note that the wavelength range by construction has to be redward of any significant damping wing absorption. 
    
    It is important to note that the actual range of the prior is not important for the reconstruction of the \lya{} line profile \citep{Greig:2016p1}. However, since we fit the IGM damping wing far redward of the line center, out to 1230\AA\ it is critical that this prior range includes sufficient information to allow for the fitting of the nearby \nv{} line. While near the \lya{} line centre, the total QSO flux is expected to be solely dominated by the \lya{} line, further redward of the line centre one can expect an increasing contribution from the \nv{} line as its line centre is approached.  Unfortunately, simultaneously reconstructing the \lya{} and \nv{} emission lines would yield uncertainties which are too large.  Thus we impose a prior just blueward of the \nv{} line.
    As this lower limit is increased from our adopted 1230\AA\ moving into the line, the spread in the allowed QSO flux will increase, broadening the recovered  constraints on  the IGM neutral fraction.

 \end{itemize}

\subsubsection{Reconstruction of ULASJ1120+0641} \label{sec:ULASJ1120}

\begin{figure*} 
	\begin{center}
		\includegraphics[trim = 0.4cm 0.3cm 0cm 0.5cm, scale = 0.50]{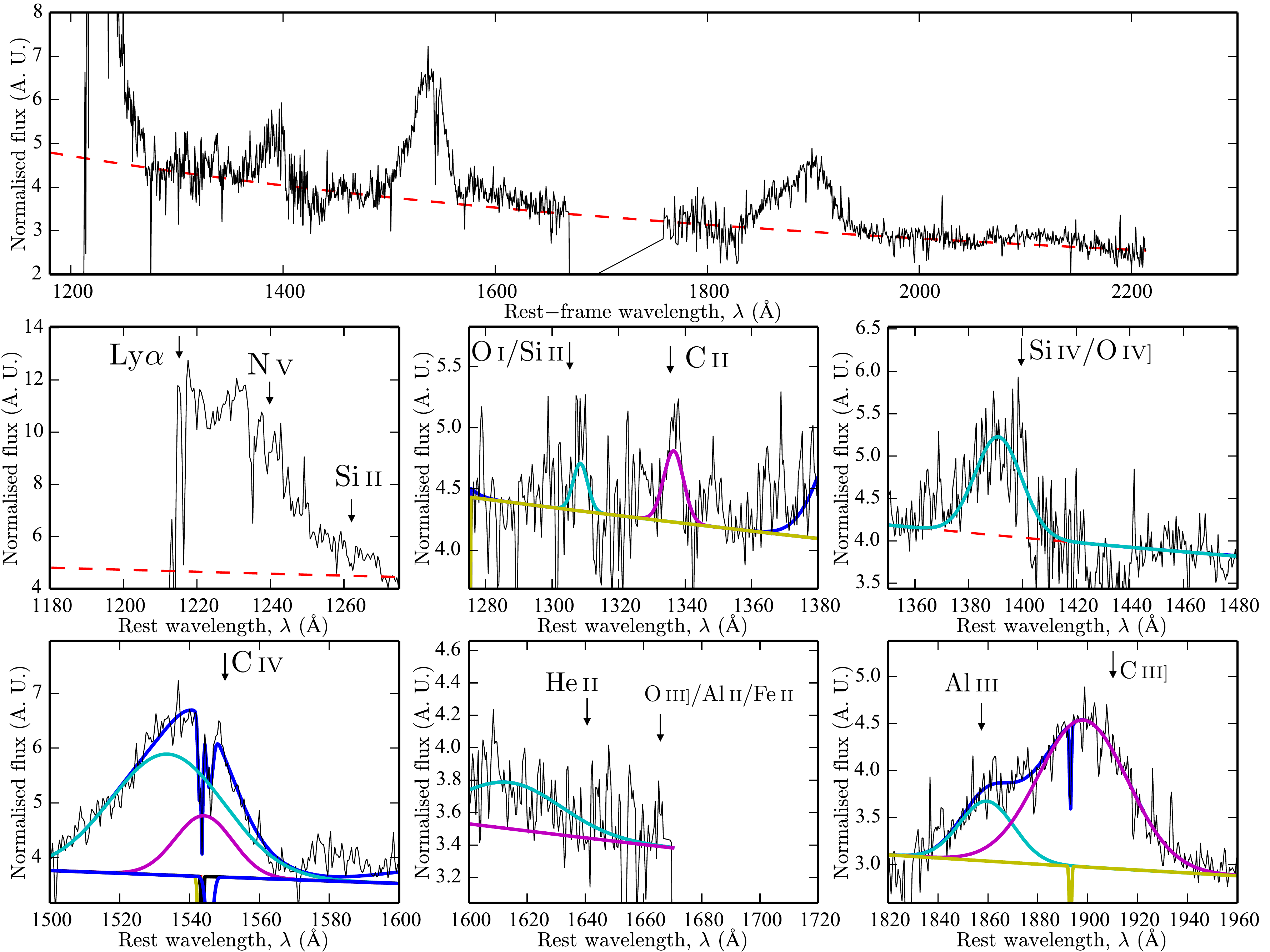}
	\end{center}
\caption[]
{A zoom-in highlighting the MCMC \qso{} fitting procedure of \citet{Greig:2016p1} applied to the rest-frame FIRE spectrum \citep{Simcoe:2012p1057}. This method includes the identification of `absorption' features (e.g.\ the bottom left and bottom right panels), which improves our ability to recover the emission line profiles. The flux is normalized to unity at 1450\AA~rest-frame (1 A. U. = $10^{-17}\,{\rm erg\,cm^{-2}\,s^{-1}\,\AA^{-1}}$). Arrows denote the systemic redshift of each line, obtained from the recovered atomic [C\,{\scriptsize II}] redshift \citep{Venemans:2012p4996}.  \textit{Top:} A single power-law continuum component fit to the QSO spectrum (red dashed curve). \textit{Middle left:} The obscured \lya{} peak profile (not-fit). \textit{Middle centre}: Two low-ionisation lines, O\,{\scriptsize I}/Si\,{\scriptsize II]} (cyan) and C\,{\scriptsize II} (magenta). \textit{Middle right:} Single component Gaussian fit to the \sioiv{} blended line complex. \textit{Bottom left:} Two-component fit to \civ{}. \textit{Bottom centre:} Low ionisation lines, He\,{\scriptsize II} (cyan; no prevalent emission, therefore the line profile is unconstrained) and O\,{\scriptsize III} (within the excised region, therefore not fit). \textit{Bottom right:} Single component fit to \ciii{} (magenta) and single Gaussian Al\,{\scriptsize III} component (cyan).}
\label{fig:Spectrum}
\end{figure*}

In Figure~\ref{fig:Spectrum}, we present the MCMC template fitting of ULASJ1120 at $\lambda > 1275$\AA.
In the top panel, the red-dashed curve corresponds to the best-fit QSO continuum, whereas in the remaining zoomed-in panels we present the best-fits to the various emission line profiles, using either a single or double component Gaussian as described above. Compared to the QSO spectra used in the construction of the covariance matrix in \citet{Greig:2016p1}, the FIRE spectrum is considerably noisier.

Immediately obvious from Figure~\ref{fig:Spectrum} is the significant blueshift observed amongst all the high-ionisation lines. In addition to the already reported strong blueshift of \civ{} \citet{Mortlock:2011p1049}, we note that the \nv{} (not fit), \sioiv{} and \ciii{} lines appear to be equally strongly blue shifted. At the same time, the low ionisation lines, O\,{\scriptsize I} and C\,{\scriptsize II} do not appear to be blue shifted at all. Furthermore, while not shown in Figure~\ref{fig:Spectrum} or observed in the \citet{Simcoe:2012p1057} FIRE spectrum, the \mgii{} line also does not have a significant blue shift \citep{Mortlock:2011p1049}. This behaviour is well known, which highlights that the physics governing the low and high ionisation lines stem from different processes or physical regions. We stress that this strong observed blueshift in all high-ionisation lines is automatically accounted for by the covariance matrix reconstruction pipeline.

Note that, in the FIRE spectrum, the \sioiv{} line is strongly affected by both night sky OH lines and telluric absorption bands. While attempts were made to correct this in the spectrum, they will still leave an imprint in the form of lower signal to noise and larger residuals. Within this work we attempt to mask out the worst of these lines, however we caution that the \sioiv{} line may still be contaminated. However, we note that the \sioiv{} emission line is the least important of the three high-ionisation lines used to reconstruct the intrinsic \lya{} line profile. Therefore, while the characterisation of the \sioiv{} line profile is likely to be contaminated, the uncertainties arising from this will not greatly impact our reconstructed \lya{} line profile.

\begin{figure*} 
	\begin{center}
		\includegraphics[trim = 0cm 0.8cm 0cm 0.5cm, scale = 0.55]{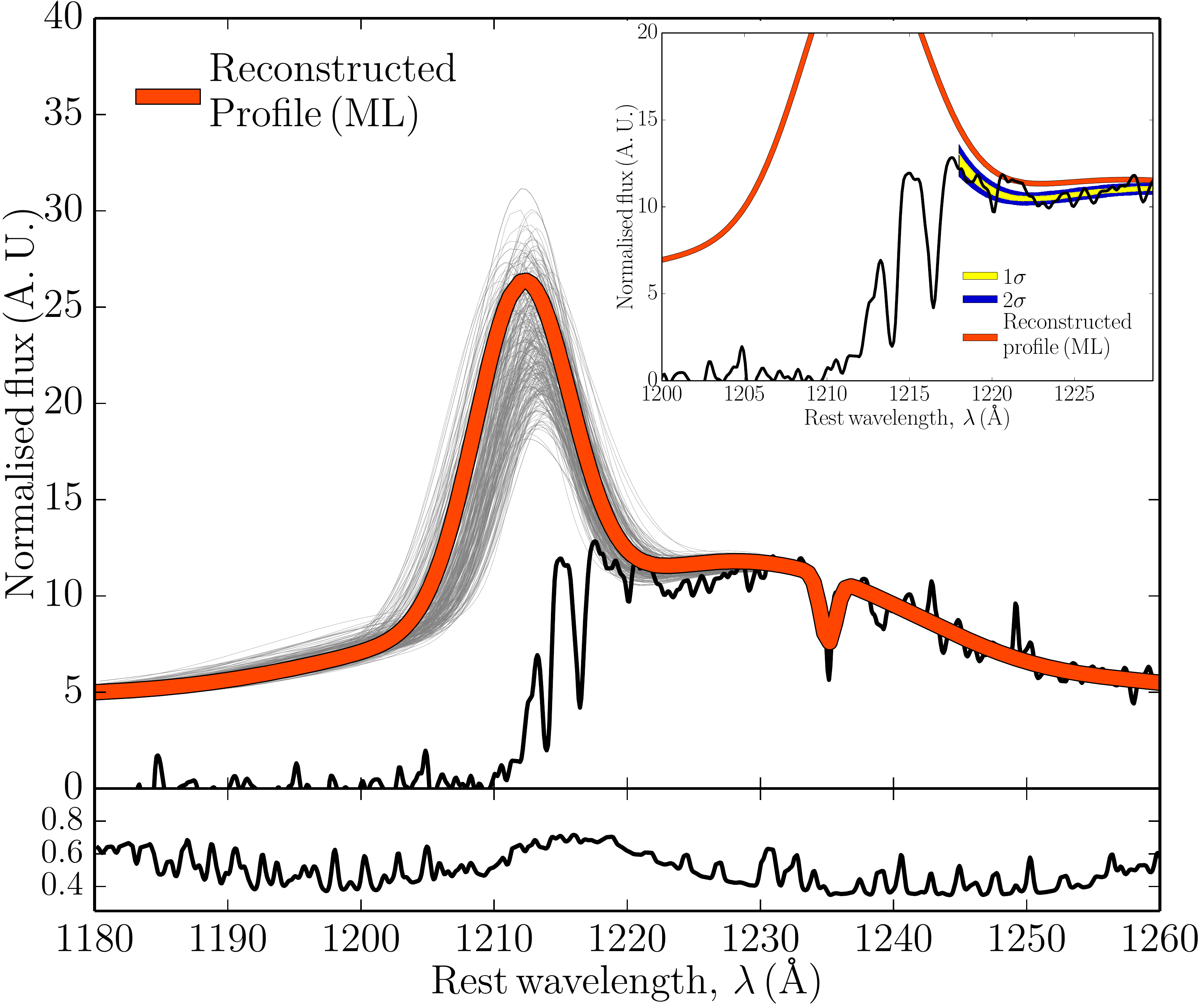}
	\end{center}
\caption[]
        {The reconstructed maximum likelihood \lya{} emission line profile (red curve; shown for visualisation purposes only) recovered from our covariance matrix method including an additional prior on the flux amplitude within the wavelength range $1230 < \lambda < 1275$\AA. The thin grey curves denote a subsample of 300 \lya{} line profiles extracted from the reconstructed six-dimensional \lya{} likelihood function (note, we sample $10^5$ profiles in our full analysis). These curves are selected to be within 68 per cent of the maximum likelihood profile, and highlight the relative scale of the errors on the reconstructed \lya{} line profile and fully encapsulate the intrinsic object-to-object variation of the QSO sample. The black curve denotes the observed spectrum of ULASJ1120, which has been resampled onto 0.1\AA\ bins purely for the purposes of this figure (the full analysis pipeline uses the unaltered spectrum as shown in Figure~\ref{fig:Spectrum}). \textit{Lower panel:} The corresponding unaltered error spectrum of ULASJ1120 averaged onto 0.1\AA\ bins. \textit{Inset:} A zoom-in around \lya{} highlighting the recovered imprint of the IGM damping wing profile (Section~\ref{sec:Constraints}). The yellow (blue) shaded region denotes the $1\sigma$ ($2\sigma$) span of total (intrinsic + damping wing absorption) flux over the fitted region.  For reference, the red curve is the maximum likelihood intrinsic profile (same as main figure).  The stretch between 1222 -- 1227 \AA\ is especially difficult to match without the aid of the IGM damping wing from an incomplete reionisation.}
\label{fig:Profile}
\end{figure*}

In Figure~\ref{fig:Profile}, we show the reconstructed intrinsic \lya{} emission line profile. The red curve is the best-fit reconstructed profile (shown for visualisation purposes only) obtained by jointly sampling the \lya{} line profile and the wavelength range $1230 < \lambda < 1275$\AA, while the black curve is the observed spectrum sampled in 0.1\AA\,bins\footnote{Note, the choice to resample the spectrum is purely for visual representation. The original, unaltered spectrum (e.g.\ Figure~\ref{fig:Spectrum}) is used within our full analysis pipeline.}. Given that our reconstruction procedure returns a six-dimensional likelihood function, in order to characterise the scale of the 68 per cent uncertainties on the reconstructed \lya{} profile we randomly sample this likelihood function to extract representative profiles. In Figure~\ref{fig:Profile} we present a subsample of 300 reconstructed \lya{} line profiles denoted by the thin grey lines which are within the 68 per cent uncertainties. These profiles highlight the relative scale of the variations in the total \lya{} line profile peak height, width and location. Note, our analysis pipeline operates directly on a large number of these sampled profiles (drawn from the full distribution), and not the maximum-likelihood reconstructed profile (red curve).

Qualitatively, the reconstructed \lya{} line profile is similar to the composite  QSO spectrum presented in \citet{Mortlock:2011p1049} and in \citet{Simcoe:2012p1057}, albeit with a slightly higher amplitude at the \lya{} line centre. A notable advantage of our approach is the statistical characterisation of the \lya{} line offset, using the covariance matrix. As shown in \citet{Bosman:2015p5005}, the location of the \lya{} line centre of the reconstructed profile is crucial for the analysis of the IGM damping wing.  Unfortunately, the blue shifts observed in ULASJ1120 exceed those found in all of the 1673 QSOs from which we construct the covariance matrix. The mean blue shift of the \civ{} ionisation line profile of ULASJ1120 (obtained from combining the narrow and broad components), $\Delta v \sim2\,500$~km/s, is a $\sim4\sigma$ outlier from our QSO sample.  Thus in the reconstruction procedure for  ULASJ1120, we have to extrapolate the Gaussian covariances.

We can test this extrapolation in two ways: (i) by testing our reconstruction approach against the most heavily blue shifted QSOs within the sample used to construct the covariance matrix and (ii) by artificially removing the strong observed blue-shift. In Appendix~\ref{sec:VelOffsetTest} we explore the former, finding little evidence for a reduction in the quality of reconstructed \lya{} profiles for the 50 most extreme objects in our sample\footnote{Note that the \civ{} blueshifts for this sample of 50 QSOs were between 1,300 - 2,000~km/s, smaller than the recovered ~2,500~km/s of ULASJ1120 (see Figure~\ref{fig:VelOffsetTest}). }.
For the latter, we take the systemic redshift to be that from the \ciii{} line (instead of our fiducial atomic [\cii{}]). As noted in Section~\ref{sec:ULASJ1120}, the \ciii{} emission line appears equally strongly blue shifted as \civ{} and the other high ionisation lines; therefore, by using the \ciii{} redshift we can artificially remove the strong \civ{} velocity offset.
After performing the \lya{} reconstruction pipeline on this artificially corrected spectrum, we recover a quantitatively similar \lya{} line profile shape, well within our 1-$\sigma$ uncertainties.
Therefore, following these two tests, we can be relatively confident in the extrapolation of these Gaussian covariances to the values found in ULASJ1120.

\subsection{The IGM damping wing during the EoR} \label{sec:IGMDampingWing}

As stated above, the observed spectrum redward of the \lya\ line centre depends on the intrinsic emission (discussed in the previous section), and the damping wing absorption from (putative) cosmic \hi{} patches along the line of sight (LoS). To statistically characterise the latter, we make use of the publicly-available Evolution of 21-cm Structure (EOS; \citealt{Mesinger:2016p1})\footnote{http://homepage.sns.it/mesinger/EOS.html} 2016 data release. These semi-numerical reionisation simulations are 1.6~Gpc on a side with a 1024$^3$ grid, and include state-of-the-art sub-grid prescriptions for inhomogeneous recombinations and photo-heating suppression of star-formation. The 2016 EOS data release corresponds to two simulation runs, with the efficiency of supernovae feedback adjusted to approximately bracket the expected EoR contribution from faint galaxies.  The two runs are:
\begin{itemize}
\item {\bf \faintgal} --  the EoR is driven by galaxies residing in haloes with masses of $10^8 \lesssim M_h/M_\odot \lesssim 10^9$, and is characterised by numerous small cosmic \hii{} regions. Hereafter we will refer to the EoR morphologies resulting from this simulation as {\bf \smallHII}.
\\
\item {\bf \brightgal} --  the EoR is instead driven by galaxies residing in haloes with masses of $M_h \sim 10^{10} M_\odot$, and is characterised by spatially more extended \hii{} structures.  We refer to these simulated EoR morphologies as {\bf \largeHII}.
\end{itemize}
We note that the \largeHII\ ionisation fields have a factor of $\sim$few--10 times more power on large scales during the EoR, compared to the \smallHII\ ionisation fields.
Although these are two opposite extremes in terms of \hii{} region sizes, the \smallHII\ EoR morphology is likely more realistic (see the discussion in \citealt{Mesinger:2016p1}), and so we use this simulation as our fiducial model.
Nevertheless, we include both extremes to explore the dependence of our results on EoR morphology. 

We construct samples of sightlines through our simulations which are terminated on one end at a halo, and then extended in a random direction through the simulation volume for a distance of 200 comoving Mpc.  Along each sightline, we sum the contributions from all encountered \hi{}\ patches to construct a composite optical depth for the damping wing (e.g. \citealt{MiraldaEscude:1998p1041}). We purposefully exclude the \hi{} contribution from pixels $\leq16$ comoving Mpc (2 physical Mpc) from the QSO host halo.  This corresponds to the minimum possible radius of the \hii{} region around ULAS1120 inferred from measurements of its near zone \citep{Mortlock:2011p1049}. We do not explicitly include the flux from the QSO in our reionisation maps.  Doing so would require assumptions about the QSO lifetime and ionising luminosity.  Neglecting the QSO ionising contribution implies that the surrounding \hii{} region could be larger than predicted by our EoR models.  For a given \avenf\, a larger surrounding \hii{} region implies a smaller integrated damping wing optical depth.  Thus if the QSO contributes in growing the surrounding \hii{} region, our modelled optical depths should be associated with even higher values of \avenf, shifting the PDFs we present below towards larger \avenf. 

At $z=7.1$, we select $10^4$ identified haloes in the mass range $6\times10^{11} < M_h/M_\odot < 3\times10^{12}$, consistent with the inferred dynamical mass of the host halo of ULASJ1120 \citep{Venemans:2012p4996}. Importantly, these higher host halo masses, made possible by the large-scale EOS simulations, are a notable improvement over previous studies \citep[e.g.][]{Bolton:2011p1063} as they better capture the bias and scatter of the QSOs locations inside reionisation fields. Though the lines of sight begin 16 Mpc from the host haloes, this bias and scatter may still be important (see e.g.\ fig.~2 of \citealt{Mesinger:2008p3610} and fig.~3 of \citealt{Mesinger10}). We use 10 LoSs per host halo, resulting in a total sample of $10^5$ synthetic IGM damping wing profiles for each sampled \avenf\ and EoR morphology (\smallHII\ and \largeHII). Since we wish to leave the IGM neutral fraction at $z=7.1$ as a free parameter, we follow the common practice of sampling ionisation fields at various redshifts corresponding to a given \avenf. That is, we use the same halo list obtained from the $z=7.1$ snapshot to define the locations for our synthetic damping wing profiles, but vary the mean IGM neutral fraction by sampling the corresponding ionisation fields obtained from different redshift outputs. Such an approach is justified as the
  ionisation fields are largely redshift independent, when compared at a fixed \avenf\ (e.g. \citealt{McQuinn07LAE, McQuinn07, SM15})

\subsection{Joint fitting of the intrinsic emission and IGM damping wing} \label{sec:Constraints}

\begin{figure*} 
	\begin{center}
		\includegraphics[trim = 0.2cm 0.6cm 0cm 0.5cm, scale = 0.68]{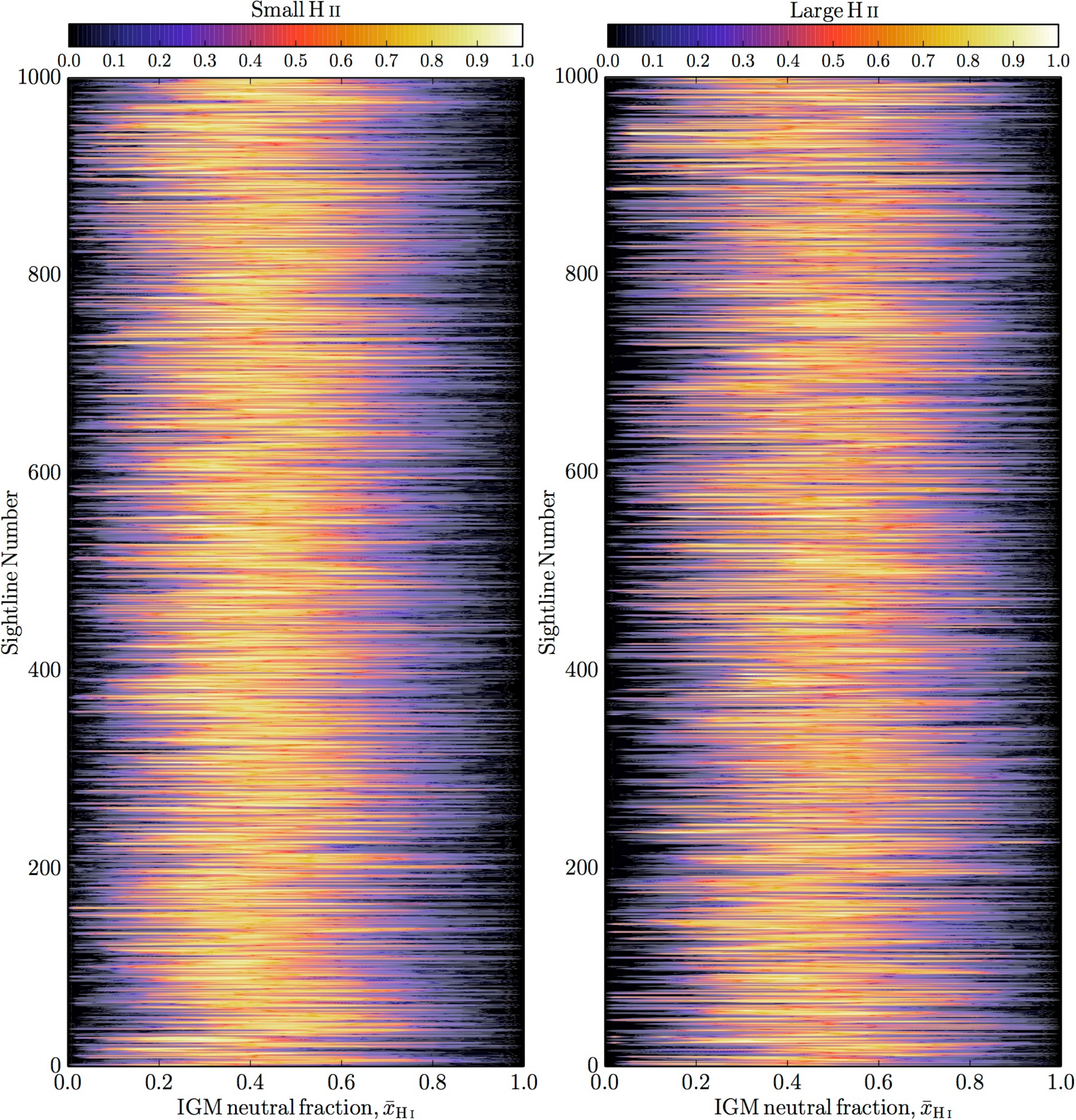}
	\end{center}
\caption[]
{1D PDFs of the IGM neutral fraction drawn from a subsample of 1000 lines of sight for each of the two EoR simulations used in our analysis \citep{Mesinger:2016p1}. Colour bars denote the amplitude of the PDFs, P($\bar{x}_{\hi{}}$). Note that the peaks of the PDFs of each sightline are normalised to unity purely to aid the visualisation. The shifting locations of the peaks per sightline are indicative of the sightline-to-sightline variation. In the left panel, the individual sightline PDFs correspond to the \smallHII\ EoR simulations (reionisation driven by faint galaxies producing small cosmic \hii{} regions) whereas the right panel corresponds to the \largeHII\ EoR simulations (reionisation driven by bright galaxies producing large cosmic \hii{} regions).  Averaging over the full sample of 10$^5$ sightlines (i.e. collapsing along the vertical direction) results in the 1D PDFs of \avenf\ shown in the following figure.}
\label{fig:Sightlines}
\end{figure*}

Having outlined the reconstruction of the intrinsic \lya{} line profile of ULASJ1120 in Section~\ref{sec:Reconstruction} and the IGM damping wing profiles in Section~\ref{sec:IGMDampingWing} we now combine them to simultaneously fit the observed spectrum.  
Our fitting procedure consists of the following steps:
\begin{enumerate}
\item The intrinsic \lya{} line profile recovered in Section~\ref{sec:ULASJ1120} is fully described by a six dimensional likelihood function (three parameters for each of the two Gaussian components), characterising the uncertainties and correlations amongst the \lya{} line profile parameters, constrained by the spectrum at $\lambda >1230$\AA.  We draw $\sim10^5$ \lya{} line profiles directly from this six dimensional likelihood.
\item Each line profile is then multiplied by each of the 10$^5$ EoR damping wing absorption profiles, resulting in a total sample of $\sim10^5 \times 10^5$ mock spectra for each value of \avenf.
\item Each mock spectrum is then compared with the observed spectrum of ULASJ1120 in the wavelength range $1218$\AA $ < \lambda < 1230$\AA\footnote{The choice of 1230\AA\ is motivated by the blue edge of the prior discussed in Section~\ref{sec:ULASJ1120}, while the choice of $1218$\AA\ is motivated by ensuring that we are sufficiently far from the influence of any infalling or local gas (e.g. \citealt{Barkana:2004p4778}), which is not modelled by our EoR simulations. Note that the circular velocity of the host halo (e.g. \citealt{Venemans:2012p4996}) corresponds to a rest frame offset of $\sim2$ \AA.  We verify that changing this range only impacts our \avenf\ constraints at the percent level by considering the following alternatives: $1220$\AA $ < \lambda < 1230$\AA, $1220$\AA $ < \lambda < 1228$\AA\ and $1218$\AA $< \lambda < 1228$\AA.}.  The quality of the fit is characterized by a ($\chi^{2}$ based) likelihood, using the observational errors of the spectrum.
\item The resulting likelihood, averaged over all $\sim10^{10}$ mock spectra, is then assigned to that particular \avenf.
\item Steps (ii)--(iv) are repeated for each trial value of \avenf. We sample the range $0.01 \leq\ $\avenf~$\leq  0.95$, with 40 (28) individual snapshots for the \smallHII\ (\largeHII) simulations (note that the EoR proceeds more rapidly in the \largeHII\ model, resulting in a coarser \avenf\ sampling for outputs at fixed redshift intervals).
  \item We normalise the resulting relative likelihoods, ending with a final 1D probability distribution function (PDF) of \avenf\ for each of the EoR morphologies.
\end{enumerate}
The above steps effectively result in the construction a 3D likelihood which is a function of: (i) \avenf; (ii) the EoR damping wing sightline; and (iii) the intrinsic emission profile.  Our final constraints on \avenf\  are obtained by marginalising over (ii) and (iii).

\section{Results} \label{sec:results}

In the inset of Figure~\ref{fig:Profile}, we present the confidence intervals on the product of the reconstructed \lya{} line profile and the synthetic IGM damping wing profiles within the fitting interval $1218$\AA\ $< \lambda < 1230$\AA. For reference, the red curve is the intrinsic \lya{} emission line profile with the maximum likelihood from our reconstruction procedure. The impact of the damping wing contribution is highlighted by the offset of the shaded regions and the red curve. In the main panel of Figure~\ref{fig:Profile}, we present a small subset (300) of recovered \lya{} line profiles to convey the relative uncertainties in the reconstruction pipeline. Note that the wavelength stretch between 1222 -- 1227 \AA\ is especially difficult to fit purely with the intrinsic profiles alone (in.  As we shall see below, we require a non-zero IGM damping wing contribution to fit the observed spectrum of ULASJ1120.

\begin{figure} 
	\begin{center}
		\includegraphics[trim = 0.2cm 0.8cm 0cm 0.5cm, scale = 0.42]{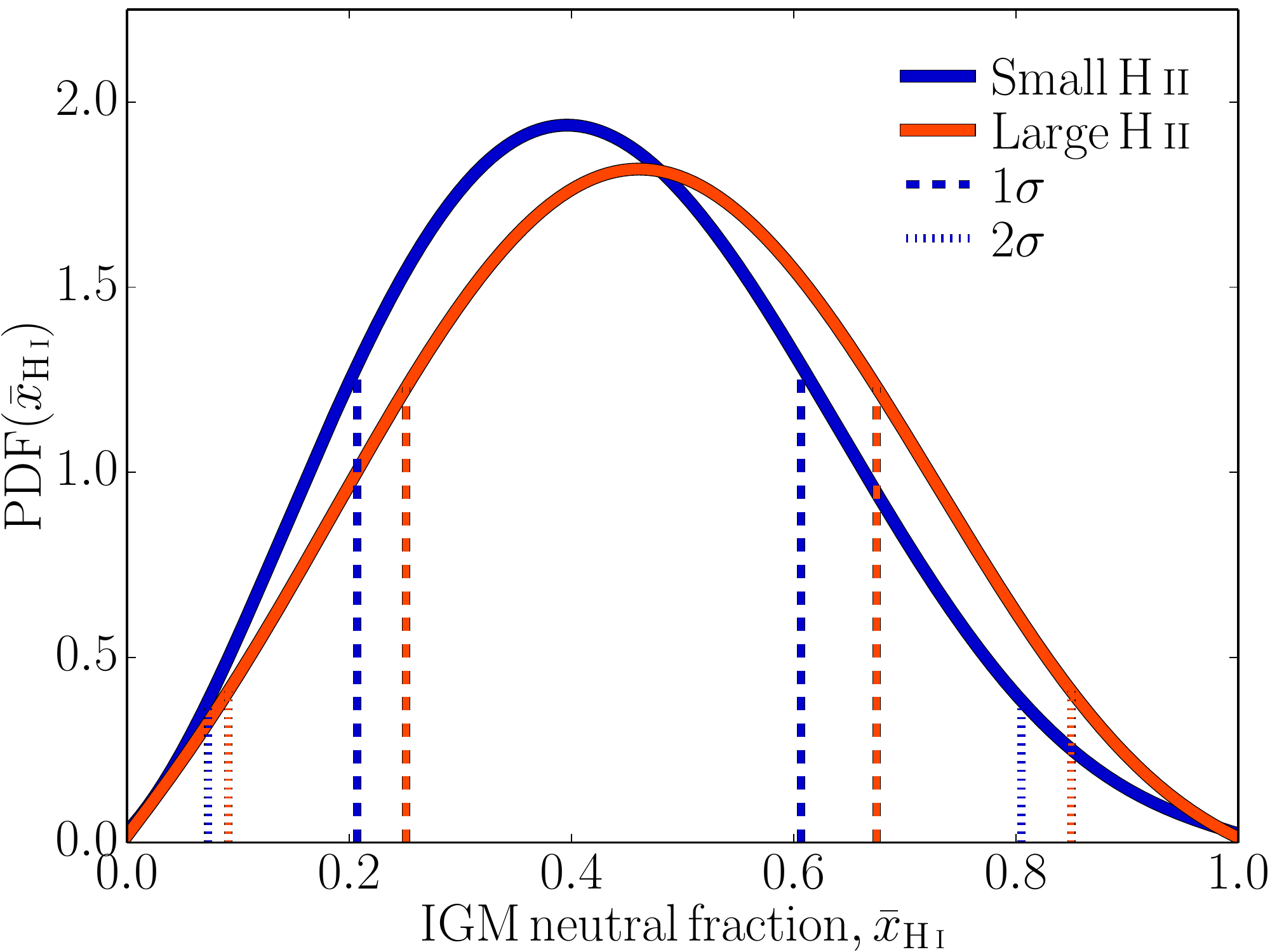}
	\end{center}
        \caption[]{PDFs of the $z=7.1$ IGM neutral fraction obtained by marginalising over all synthetic IGM damping wing absorption profiles and reconstructed intrinsic \lya{} emission line profiles. The red curve corresponds to the \largeHII\ simulations (right panel of Figure~\ref{fig:Sightlines}) whereas the blue curve corresponds to the \smallHII\ simulation (left panel of Figure~\ref{fig:Sightlines}). Dashed (dotted) curves correspond to the 1 (2$\sigma$) constraints on \avenf\ for the respective morphologies. Both simulations recover consistent results, favouring a strong damping wing imprint from a significantly neutral IGM ($\bar{x}_{\hi{}}\sim0.4$).}
\label{fig:PDFs}
\end{figure}

Before presenting our final constraints on \avenf, we showcase the EoR sightline-to-sightline scatter in Figure \ref{fig:Sightlines}.  For each of a randomly selected subsample of 1000 sightlines shown in the figure, we average over the full distribution of the reconstructed \lya{} intrinsic profiles, in order to generate a \avenf\ PDF for that sightline. Collapsing (marginalising) over the vertical direction (sightline number) for the entire sample of $10^5$ LoSs recovers the full 1D marginalised PDF (step (vi) of Section~\ref{sec:Constraints}; see Figure~\ref{fig:PDFs}). Sightlines extracted from the \smallHII\ and \largeHII\ simulations appear on the left and right, respectively.

On average, we recover a similar range for the preferred IGM neutral fraction for both EoR morphologies. However, there is significant sightline-to-sightline variation, shifting the peaks of the \avenf\ distributions by tens of per cent. This highlights the importance of sampling a large number of IGM damping wing profiles. Our full sample consists of $10^5$ sightlines through inhomogeneous reionisation, compared with 100 sightlines through homogeneous reionisation in the preliminary studies of \citet{Bolton:2011p1063} and \citet{Keating:2015p5004}.

Finally, in Figure~\ref{fig:PDFs} we present the main result of this work: 1D PDFs of \avenf \footnote{
  
In Appendix~\ref{sec:ContinuumExploration}, we explore the impact of potential errors in the flux calibration of ULASJ1120, and how this could impact the overall constraints on the IGM neutral fraction. By including a conservative 10~per cent error on the QSO continuum we find, for the \smallHII\ model, constraints consistent with the main results of this paper.
}.  These are obtained by marginalising over all combinations of reconstructed intrinsic \lya{} line profiles and synthetic IGM damping wing sightlines for a given \avenf. The blue curve corresponds to the \smallHII\ EoR morphology, while the red curve corresponds to the \largeHII\ EoR morphology. Dotted (dashed) curves correspond to the 1 (2$\sigma$) constraints on \avenf\ for the respective morphologies:
\begin{itemize}
\item {\bf \smallHII}; $\bar{x}_{\hi{}} = 0.40\substack{+0.21 \\ -0.19}$ ($1\sigma$) and $0.40\substack{+0.41 \\ -0.32}$ ($2\sigma$) \\
\item {\bf \largeHII}; $\bar{x}_{\hi{}}  = 0.46\substack{+0.21 \\ -0.21}$ ($1\sigma$) and $0.46\substack{+0.39 \\ -0.37}$ ($2\sigma$). \\
\end{itemize}
As mentioned earlier (and discussed in \citealt{Mesinger:2016p1}), the \smallHII\ model is likely more accurate and we adopt it as our fiducial constraint.  We note that the constraints on \avenf\ are very similar for both \smallHII\ and \largeHII.  This indicates that the damping wing imprint is not very sensitive to how the cosmic neutral patches are distributed at a fixed value of \avenf.

  From the recovered PDF of the IGM neutral fraction we can strongly infer the presence of a damping wing signature, with no IGM damping wing contribution being inconsistent at $>2\sigma$.
  This  is despite the fact that at first glance in Figure~\ref{fig:Profile} the reconstructed \lya{} profiles (without damping wing absorption) appear to be consistent with the observed spectrum of ULASJ1120 in the lower end of the 1218--1221\AA\ wavelength range used in our analysis.
  In Appendix~\ref{sec:FittingRange}, we quantify this
  in detail by performing our full analysis pipeline on smaller, equally spaced 3\AA\ sub-regions.
  The recovered constraints from the sub regions are consistent with each other at 1$\sigma$, with the strongest evidence for a damping wing coming from the 1222-1227\AA\ region of the observed spectrum.

  \textit{How do these constraints tie into the existing picture of reionisation?} The best available, model independent constraints on the end stages of reionisation are obtained from the dark fraction of pixels in QSO spectra \citep{McGreer:2015p3668}, which imply the IGM neutral fraction to be \avenf\ $\lesssim 0.11$ (1$\sigma$). Therefore the \avenf\ $= 0.40\substack{+0.21 \\ -0.19}$ constraints obtained here, imply an evolution of $\Delta$\avenf\ $\sim$ 0.1--0.5 over the redshift interval $z \approx 7\rightarrow 6$.  This is  consistent with the currently available EoR constraints (see \citealt{Greig:2016EORhist} and references therein).

\section{Could the damping wing come from a damped Lyman alpha system?} \label{sec:Discussion}

In this work we found evidence of a damping wing imprint on the red side of the \lya\ line of ULASJ1120, and quantified the corresponding constraints on the IGM neutral fraction.  However, a damping wing could instead be produced by an intervening damped \lya{} system, at least in principle.  Indeed most high-$z$ gamma-ray bursts (GRBs) show evidence of a DLA in their spectra \citep[e.g.][]{Chornock:2013p4098,Chornock:2014p4292,Totani:2014p4293}. However, the GRB DLAs are associated with the GRB host galaxy, while a DLA in the spectra of ULASJ1120 would have to be at least 16~Mpc away from the QSO host galaxy. As has been pointed out previously, finding such an object in a random skewer through the IGM is highly unlikely. To mimic our results, the required column density of the DLA would have to be ${\rm log_{10}}(N_{\hi{}}/{\rm cm^{-2}}) > 20 - 21$ (see also \citealt{Simcoe:2012p1057,Schroeder:2013p919}). Such systems are extremely rare. Following \citet{Schroeder:2013p919}, we note that an extrapolation of the incidence rate of DLAs \citep[][]{Prochaska:2009p3339,Songaila:2010p3348} to $z\sim7$ implies an abundance of $\lsim 0.05$ DLAs per unit redshift with comparable column densities (see fig. 8 in \citealt{Songaila:2010p3348}). Thus only $\lsim0.3$ per cent of IGM segments with a length corresponding to the ULASJ1120 near zone size ($\Delta z_{\rm NZ} \sim 0.05$) would contain a DLA.  We also note that \citet{Bolton:2011p1063} found only $\sim5$\% of their mock sightlines contained a DLA, even using a self-shielding prescription which notably overestimates their abundances \citep{Rahmati:2013p1,Keating:2014p1,Mesinger:2015p1584}. Thus, the a-priori presence of a DLA can approximately be excluded at $\gsim$~2--3$\sigma$ on the basis of the required column density alone.

Even more damning is the fact that there is no evidence of associated metal line absorption in ULASJ1120.  Therefore a putative DLA would have to be atypically pristine, with a metallicity of $Z \lsim10^{-4} Z_\odot$ which is inconsistent with every other DLA observation \citep[e.g.][]{Simcoe:2012p1057,Cooke:2015}. We therefore conclude that the damping wing absorption seen in ULASJ1120 is highly unlikely to originate from a DLA.

\section{Conclusion} \label{sec:Conclusion}

In this work, we obtain constraints on the $z=7.1$ IGM neutral fraction by isolating its damping wing absorption in the spectrum of QSO ULASJ1120.
We use a state-of-the-art Bayesian framework which for the first time is able to jointly sample both: (i) the uncertainty in the QSO intrinsic emission; and (ii) the EoR sightline-to-sightline variation.  For (i), we use a covariance matrix of emission line properties \citep{Greig:2016p1} to reconstruct the intrinsic \lya\ line profile.  For (ii), we use the latest, large-scale simulations of patchy reionisation (Mesinger et al. 2016). After marginalising over (i) and (ii), we obtain robust constraints on the IGM neutral fraction.  For our fiducial reionisation model, these are: $\bar{x}_{\hi{}} = 0.40\substack{+0.21 \\ -0.19}$ ($1\sigma$) and $0.40\substack{+0.41 \\ -0.32}$ ($2\sigma$).  We note that the constraints are very insensitive to the EoR model, at a fixed global neutral fraction (see Figure \ref{fig:PDFs}).

Our results correspond to the first measurement of the ionisation state of the IGM at $z\sim7$, with a well-defined confidence range (as opposed to upper/lower limits).  They are consistent with the latest {\it Planck} measurements of the Thompson scattering optical depth, which independently appeared as this work was nearing completion \citep{Collaboration:2016p5913}.

The framework we developed can easily be applied to future QSO observations.  Moreover, the analysis can be extended to incorporate the transmission statistics in the QSO near zone (blueward of \lya).  This would introduce additional uncertainties, but would allow the analysis to be extended to other bright $z\sim6$--7 QSOs which have a much larger near zone, and thus a correspondingly weaker damping wing imprint on the red side of the line \citep{Schroeder:2013p919}.  We defer this to future work.

\section*{Acknowledgements}
We thank Ian McGreer for providing helpful comments on a draft version of this manuscript. Additionally, we thank the anonymous referee for their detailed and helpful suggestions which have improved the quality of this work.
AM and BG acknowledge funding support from the European Research Council (ERC) under the European Union's Horizon 2020 research and innovation programme (grant agreement No 638809 -- AIDA -- PI: AM). ZH is supported by NASA grant NNX15AB19G. RS is supported by NSF grant AST-1109915.

\bibliography{ms}

\appendix

\section[]{Extrapolating our model to ULASJ1120} \label{sec:VelOffsetTest}

In this section, we explore the validity of extrapolating the Gaussian covariances between various line profile parameters recovered from \citet{Greig:2016p1} in order to perform the \lya{} profile reconstruction of ULASJ1120. The necessity of performing this extrapolation arises due to ULASJ1120's extremely large \civ{} blueshift, which is larger than any of the 1673 `good' QSOs originally considered for the construction of the covariance matrix in \citet{Greig:2016p1}. Owing to the rarity of these extremely blue-shifted sources, rather than searching for a sufficient statistical sample of similar sources in the BOSS database, instead we investigate a subsample of the 50 most blueshifted QSOs within our existing QSO sample.

\begin{figure} 
	\begin{center}
		\includegraphics[trim = 0.5cm 0.4cm 0cm 0.5cm, scale = 0.58]{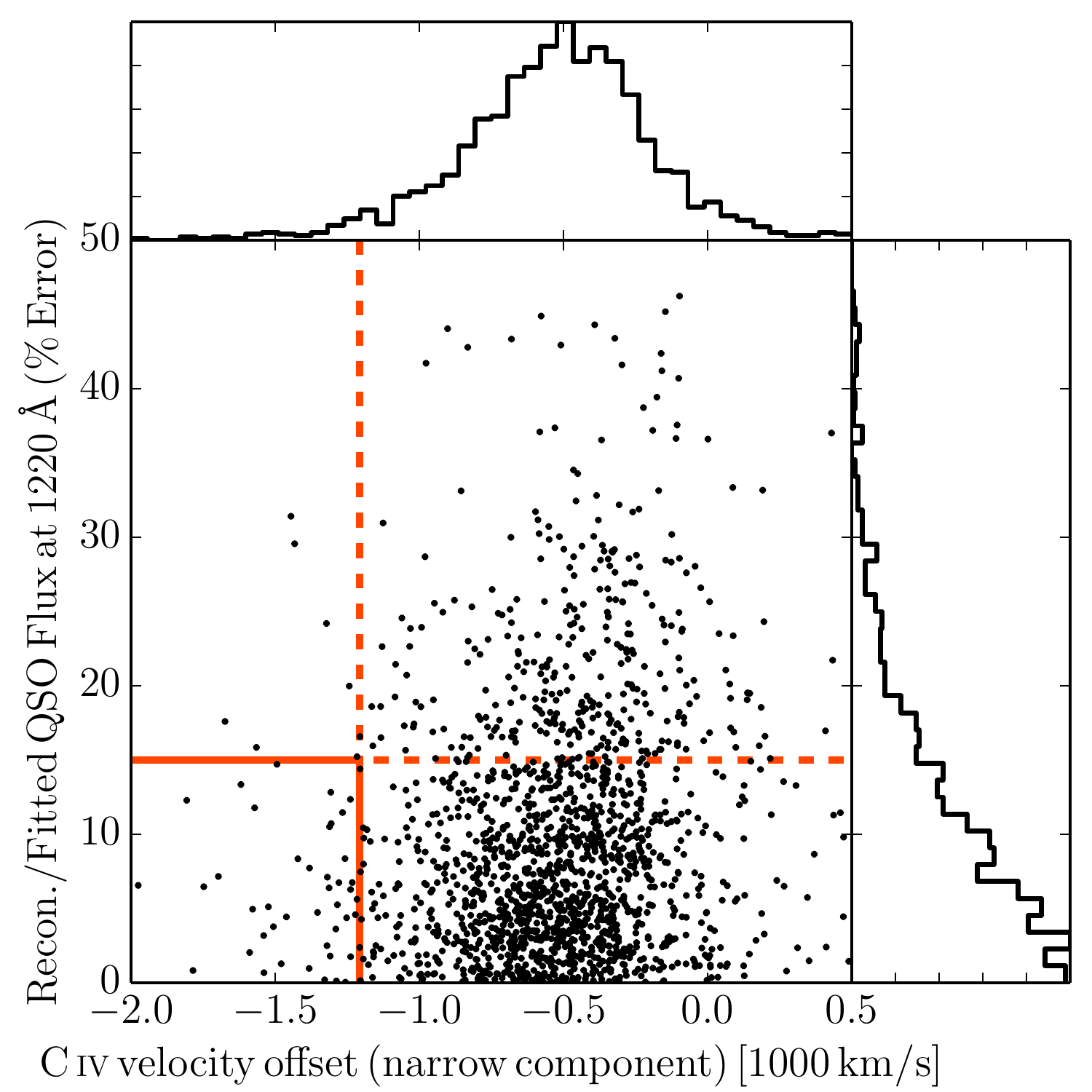}
	\end{center}
        \caption[]{The \civ{} narrow component velocity offset against the percentage error on the reconstructed QSO flux compared to the original fit QSO flux at 1220\AA\ for the 1673 QSOs within the `good' sample at $2.1 \lesssim z \lesssim 2.5$ from \citep{Greig:2016p1}. The vertical red dashed line denotes the 50 most heavily blue shifted QSOs (leftward of the line), while the red horizontal dashed line corresponds to a 15~per cent error in the reconstructed QSO flux compared to the original QSO flux. The solid red box highlights which of the 50 QSOs with the largest blueshift have reconstructed \lya{} profiles within 15 per cent of the original QSO flux.}
\label{fig:VelOffsetTest}
\end{figure}

In Figure~\ref{fig:VelOffsetTest} we present the recovered \civ{} velocity offset of the narrow line component (double Gaussian line profile) against the ratio of the reconstructed QSO flux compared to the original QSO flux at 1220\AA. In \citet{Greig:2016p1} we use this ratio as a metric to define how well the reconstruction pipeline performs. In this work, we distinguished a good characterisation of the reconstructed \lya{} profile being when this was ratio was within 15~per cent\footnote{This choice of 15~per cent was an arbitrary definition.} of the original fit to the QSO flux. As such, the horizontal red dashed line in Figure~\ref{fig:FittingRange} denotes this 15~per cent limit. The vertical red dashed line corresponds to the smallest \civ{} velocity offset of the 50 most blueshifted QSOs. The solid red box encompasses the QSOs which match both these criteria. 

For the total sample of 1673 QSOs, at 1220\AA, \citet{Greig:2016p1} found that the QSO flux could be recovered to within 15~per cent for 90 per cent of all QSOs. Applying this same criteria to the 50 most blue-shifted QSOs, we recover the QSO flux to within 15~per cent for 88 per cent of this sample. By recovering similar fractions for the 50 most blueshifted QSOs compared to the full sample (88 compared with 90 per cent), it is clear that there is no obvious decrease in performance of the reconstruction procedure with increasingly larger \civ{} blueshifts. Furthermore, Figure~\ref{fig:VelOffsetTest} shows no correlation between the quality of reconstruction and the \civ{} velocity offset. Finally, four of the five most extremely blueshifted QSOs in our sample are recovered to less than 10~per cent of the original QSO flux at 1220\AA. Therefore, given that this reconstruction procedure performs equally well irrespective of the \civ{} blueshift, we find that it is reasonable to extrapolate the covariance matrix to the extreme \civ{} blueshift of ULASJ1120.

\section[]{Impact of continuum errors owing to incorrect flux calibration} \label{sec:ContinuumExploration}

In this work, our observed spectrum of ULASJ1120 was obtained from the FIRE infrared spectrometer on the Magellan/Baade telescope. This data was measured using a narrow slit echelle spectrum, which are difficult to accurately flux calibrate.
In this section, we explore what impact any potential flux calibration errors may have had on our inferred constraints on the IGM neutral fraction.

\begin{figure} 
	\begin{center}
		\includegraphics[trim = 0.2cm 0.8cm 0cm 0.5cm, scale = 0.42]{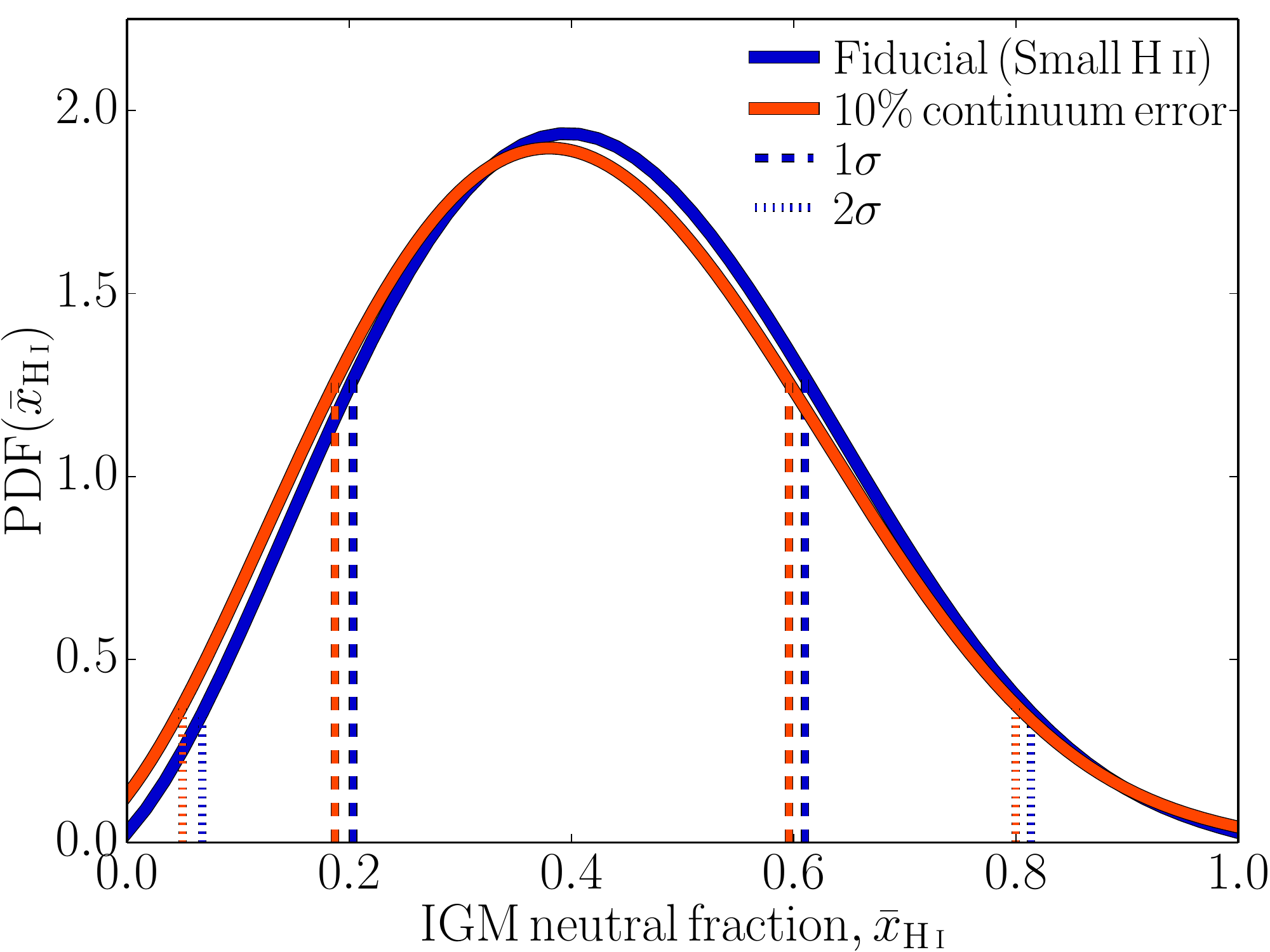}
	\end{center}
        \caption[]{PDFs of the $z=7.1$ IGM neutral fraction from the \smallHII\ simulation including an additional uncertainty to mimic potential flux calibration errors. The blue curve corresponds to our fiducial PDF (blue curve in Figure~\ref{fig:PDFs}) while the red curve has an additional 10~per cent error added to the QSO continuum which is folded into our full analysis pipeline. Dashed and dotted curves correspond to the 1 and 2$\sigma$ constraints on the IGM neutral fraction.}
\label{fig:ContError}
\end{figure}

For this, we consider a constant 10~per cent error on the fitted QSO continuum power-law from ULASJ1120 (obtained from our $>1275$\AA\ fit) over the full 1218--1230\AA\ damping wing fitting region. Note that in our reconstruction procedure it is non trivial to include a direct uncertainty on the total flux amplitude. However, within this fitting region, the amplitude of the continuum constitutes $\sim50$~per cent of the total flux, therefore a 10 per cent error on the continuum roughly equates to a error on the total flux amplitude of $\sim5$ per cent. We then repeat our full analysis pipeline outlined in Section~\ref{sec:results} fitting for the IGM damping wing imprint, adding in this 10~per cent error in quadrature. In Figure~\ref{fig:ContError}, we present the 1D PDFs of the IGM neutral fraction of the \smallHII\ synthetic damping wing profiles. The blue curve corresponds to our fiducial constraints on the IGM neutral fraction (Figure~\ref{fig:PDFs}) whereas the red curve highlights the impact of the 10~per cent error on the QSO continuum, mimicking a flux calibration error.

From Figure~\ref{fig:ContError} it is evident that the inclusion of this additional source of error on the QSO continuum has very minimal impact on our constraints on the IGM neutral fraction. By including this error, we recover an IGM neutral fraction of $\bar{x}_{\hi{}} = 0.38\substack{+0.22 \\ -0.19}$ ($1\sigma$). This is essentially equivalent to our fiducial constraints of $\bar{x}_{\hi{}} = 0.40\substack{+0.21 \\ -0.19}$ ($1\sigma$), highlighting that our results are much more sensitive to the \lya{} line recovery and the associated uncertainties, than the continuum.

This marginal reduction in the IGM neutral fraction arises from effectively broadening the distribution of reconstructed \lya{} profiles which are capable of matching the observed spectrum (i.e.\ lowering the QSO continuum level of the reconstructed profiles, requiring smaller IGM neutral fractions to match ULASJ1120). This effect is most evident by the rising amplitude in the tail of the PDF near $\bar{x}_{\hi{}} = 0$. Furthermore, as one would anticipate, by broadening the total errors applied to the recovery of the damping wing profile, the PDF of the IGM neutral fraction including this 10~per cent error, is marginally broader than the fiducial neutral fraction distribution. Importantly, given the very minor differences, we can confidently conclude that our IGM neutral fraction constraints are not strongly impacted by any potential errors arising from problems in the flux calibration of FIRE spectrum of ULASJ1120.

\section[]{Exploring various fitting ranges} \label{sec:FittingRange}

In Section~\ref{sec:results} we presented our constraints on the IGM neutral fraction by fitting for a damping wing signature between 1218--1230\AA\ in the observed spectrum of ULASJ1120.
In this section, we quantify how the different sub-regions within this wavelength range impact the final result.
To test this, we break our fiducial fitting range into four, 3\AA\ chunks and recover the inferred IGM neutral fraction for each 3\AA\ region independently.

\begin{figure} 
	\begin{center}
		\includegraphics[trim = 0.2cm 0.8cm 0cm 0.5cm, scale = 0.42]{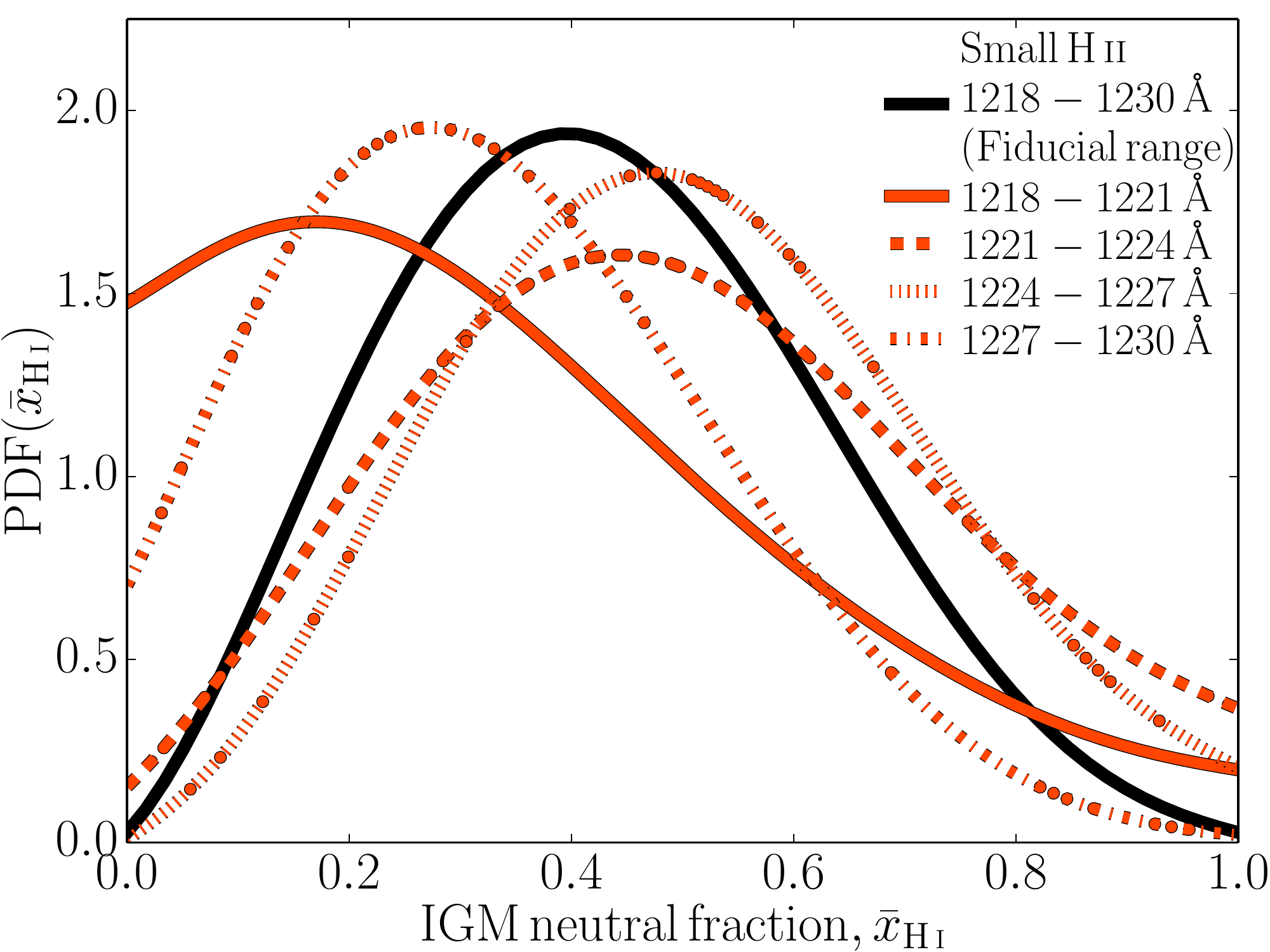}
	\end{center}
        \caption[]{PDFs of the $z=7.1$ IGM neutral fraction obtained by marginalising over all synthetic IGM damping wing absorption profiles from the \smallHII\ simulation and reconstructed intrinsic \lya{} emission line profiles for various selected fitting intervals. The solid black curve corresponds to our full fiducial fitting interval, 1218--1230\AA\ (blue curve in Figure~\ref{fig:PDFs}). The red solid, dashed, dotted and dot-dashed curves correspond to four separate 3\AA\ fitting intervals: 1218--1221, 1221--1224, 1224--1227 and 1227-1230\AA, respectively. }
\label{fig:FittingRange}
\end{figure}

In Figure~\ref{fig:FittingRange}, we present the 1D PDFs of the IGM neutral fraction at $z=7.1$ recovered from the \smallHII\ simulation, for the various 3\AA\ intervals. The black curves correspond to our fiducial 1218--1230\AA\ fitting interval, whereas the red solid, dashed, dotted and dot-dashed curves correspond to the 1218--1221, 1221--1224, 1224--1227 and 1227-1230\AA\ sub-ranges, respectively. In Table~\ref{tab:Sum}, we summarise the recovered IGM neutral fractions and associated 1 and 2$\sigma$ errors for each of the fitting intervals considered.

\begin{table}
\begin{tabular}{@{}lccc}
\hline
Fitting region (\smallHII) & \avenf (1$\sigma$) & \avenf (2$\sigma$) \\
\hline
\vspace{0.8mm}
1218--1230 \AA (fiducial) & $0.40\substack{+0.21 \\ -0.19}$ & $0.40\substack{+0.41 \\ -0.32}$\\
\vspace{0.8mm}
1218--1221 \AA  & $0.17$ ($\leq$ 0.44) & $0.17$ ($\leq$ 0.82)\\
\vspace{0.8mm}
1221--1224 \AA  & $0.44\substack{+0.27 \\ -0.22}$ & $0.44\substack{+0.50 \\ -0.36}$\\
\vspace{0.8mm}
1224--1227 \AA  & $0.47\substack{+0.23 \\ -0.20}$ & $0.47\substack{+0.42 \\ -0.35}$\\
\vspace{0.8mm}
1227--1230 \AA  & $0.28\substack{+0.21 \\ -0.19}$ & $0.28$ ($\leq$ 0.70)\\
\hline
\end{tabular}
\caption{Tabulated values of the recovered IGM neutral fraction at $z=7.1$ for the \smallHII\ simulations when considering four smaller, equally spaced 3\,\AA\ regions within our fiducial fitting range of 1218--1230\AA.}
\label{tab:Sum}
\end{table} 

Importantly, in each 3\AA\ interval there is always a clear preference for a damping wing imprint, implying we have not artificially biased our results by selecting a specific wavelength range to \textit{a priori} require a damping wing signature. Even in the 1218--1221\AA\ region, where from Figure~\ref{fig:Profile} one might anticipate we would not require a strong damping wing signature (reconstructed \lya{} profiles appear to pass through the observed spectrum), the highest likelihood is from an IGM neutral fraction of \avenf\ $\sim0.17$, although a fully ionized universe is consistent at 1$\sigma$. This interval, however, returns the largest uncertainties on the IGM neutral fraction. This is purely driven by the breadth in allowed QSO fluxes from the reconstructed \lya{} profiles, due to the fact that this region is furthest from the edge of the applied prior in the reconstruction process (1230\AA).

In contrast, the 1222--1227\AA\ region of ULASJ1120 is particularly difficult to fit without requiring higher IGM neutral fractions. This behaviour is quantified by the two adjacent 3\AA\ intervals, 1221--1224\AA\ and 1224--1227\AA. Here, larger IGM neutral fractions of $\bar{x}_{\hi{}} = 0.44$ and 0.47 are preferred.

Finally, we note that all of the sub-regions are consistent with one another at 1 $\sigma$.  Thus there is no obviously spurious spectral feature biasing our fiducial results in one direction or another.

\end{document}